\documentclass[fleqn,usenatbib]{mnras}

\usepackage{newtxtext,newtxmath}
\usepackage[font=small]{caption}
\usepackage[T1]{fontenc}

\DeclareRobustCommand{\VAN}[3]{#2}
\let\VANthebibliography\thebibliography
\def\thebibliography{\DeclareRobustCommand{\VAN}[3]{##3}\VANthebibliography}

\usepackage{graphicx}	% Including Fig. files
\usepackage{amsmath}	% Advanced maths commands
\usepackage[export]{adjustbox}
\usepackage{subcaption}

\usepackage{float}
\usepackage{color}
\usepackage{comment}

\title[Stacked X-ray clusters]{X-ray surface brightness and gas density profiles of galaxy clusters up to $3\times R_{\rm 500c}$ with SRG/eROSITA}

\author[Lyskova et al.]{N.~Lyskova,$^{1}$ E.~Churazov,$^{1,2}$ I.I.~Khabibullin,$^{3,2,1}$ R.~Burenin,$^{1}$  A.A.~Starobinsky,$^{4}$
R.~Sunyaev$^{1,2}$
\\
\\
$^1$~Space Research Institute (IKI), Profsoyuznaya 84/32, Moscow 117997, Russia \\
$^2$~Max Planck Institute for Astrophysics, Karl-Schwarzschild-Str. 1, D-85741 Garching, Germany  \\
$^3$~Universitäts-Sternwarte, Fakultät für Physik, Ludwig-Maximilians-Universität München, Scheinerstr.1, 81679 München, Germany \\
$^4$~L.D.~Landau Institute for Theoretical Physics RAS, Chernogolovka, Moscow region 142432, Russia \\
\\
}

\date{Accepted XXX. Received YYY; in original form ZZZ}

\pubyear{2015}

\begin{document}
\label{firstpage}
\pagerange{\pageref{firstpage}--\pageref{lastpage}}
\maketitle

% Abstract of the paper
\begin{abstract}
Using the data of the SRG/eROSITA all-sky survey, we stacked  
a sample of $\sim 40$ galaxy cluster images in the 0.3--2.3 keV band, covering the radial range up to $10\times R_{\rm 500c}$. The excess emission on top of the Galactic and extragalactic X-ray backgrounds and foregrounds is detected up to $\sim 3\times R_{\rm 500c}$. At these distances, the surface brightness of the stacked image drops below $\sim 1$\% of the background.  The density profile reconstructed from the X-ray surface brightness profile agrees well (within $\sim30$\%) with the mean gas profile found in numerical simulations, which predict the local gas overdensity of $\sim$ 20--30 at $3\times R_{\rm 500c}$ and the gas fraction close to the universal value of $\frac{\Omega_b}{\Omega_m}\approx 0.15$  in the standard $\Lambda$CDM model. Taking at face value, this agreement suggests that up to $\sim 3\times R_{\rm 500c}$ the X-ray signal is not strongly boosted by the gas clumpiness, although a scenario with a moderately inhomogeneous gas cannot be excluded.  A comparison of the derived gas density profile with the electron pressure profile based on the SZ measurements suggests that by $r\sim 3\times R_{\rm 500c}$ the gas temperature drops by a factor of $\sim$ 4--5 below the characteristic temperature of a typical cluster in the sample within $R_{\rm 500c}$, while the entropy keeps growing up to this distance. Better constraints on the gas properties just beyond $3\times R_{\rm 500c}$ should be possible with a sample larger than used for this pilot study.
%$R_{\rm ta}$
\end{abstract}

% Select between one and six entries from the list of approved keywords.
% Don't make up new ones.

%%%%%%%%%%%%%%%%%%%%%%%%%%%%%%%%%%%%%%%%%%%%%%%%%%

%%%%%%%%%%%%%%%%% BODY OF PAPER %%%%%%%%%%%%%%%%%%

\section{Introduction}

%\ec{ADS library (shared), "stacking-2023"}
%\ec{Density profile from Pratt+ - we can compare with them }
%\verb|https://ui.adsabs.harvard.edu/abs/2022A%26A...665A..24P/abstract| 

%\ec{after correcting for their mistake in normalization}
%\verb| https://ui.adsabs.harvard.edu/abs/2023A%26A...669C...2P/abstract|

%\ec{INTRO: NOT YET DONE}

Clusters of galaxies represent the high-mass end of the virialized halos in the present-day Universe \citep[see][for a review]{2012ARA&A..50..353K}.
To the first approximation, their properties depend on their mass and redshift. 
In particular, their radial profiles of total and gas densities and gas temperature or entropy profiles,  
once properly scaled, are expected to be approximately universal \citep[e.g. ][among others]{1998ApJ...503...77M,2005RvMP...77..207V, 2006ApJ...640..691V, 2010A&A...517A..92A, 2012MNRAS.427L..45W, 2012A&A...541A..57E, 2013A&A...551A..22E, 2019A&A...621A..41G}
although the details of the formation history of individual objects lead to inevitable deviations. 
From this perspective, averaged (stacked) profiles are useful since these deviations might be suppressed. 
The X-ray emission from massive clusters has been routinely detected up to $R_{\rm 500c}$\footnote{Throughout the paper, the halo radius $R_{\rm \Delta}$ is defined  to
enclose a fixed overdensity $\Delta$ with respect to either the critical $\rho_c$
or mean density $\rho_m$ of the Universe. The halo mass is then defined as $M_{\rm \Delta}  = 4/3 \pi R^3_{\rm \Delta} \cdot \Delta\cdot \rho_{\rm ref}$ where $\rho_{\rm ref}$ is either $\rho_m$ or $\rho_c$.}  or even up to $\approx 1.5-2 R_{\rm 500c}$ in individual cases  \citep[see, e.g., ][for reviews]{2013SSRv..177..195R,2019SSRv..215....7W, 2022hxga.book...13W}. 
The same is true for SZ signal from clusters \citep[e.g.][]{2010ApJ...716.1118P, 2013A&A...550A.131P}.
Combining the data from many clusters helps to extend the radial range probed. To this end, the data of the \textit{SRG}/eROSITA all-sky survey \citep[][]{2020Natur.588..227P, 2021A&A...656A.132S,2021A&A...647A...1P} are especially valuable since they provide uniform coverage of the regions around clusters with the same sensitivity and spatial resolution. Here we perform a pilot study of stacking a small sample of clusters using eROSITA data.

Throughout the paper, we adopted the $\Lambda$ cold dark matter cosmology with $\Omega_{\rm M} = 0.3, \Omega_{\rm \Lambda} =0.7$ and $H_0 = 70$ km s$^{-1}$ Mpc$^{-1}$. 
%The halo radius $R_{\rm \Delta}$ is defined  to enclose a fixed overdensity $\Delta$ with respect to either the critical $\rho_c$ or mean density $\rho_m$ of the Universe. The halo mass is then defined as $M_{\rm \Delta}  = 4/3 \pi R^3_{\rm \Delta} \cdot \Delta\cdot \rho_{\rm ref}$ where $\rho_{\rm ref}$ is either $\rho_m$ or $\rho_c$.

%The outer cluster regions remain  relatively unexplored, bearing signatures of  ongoing accretion processes, possible departures from virialization and hydrostatic equilibrium (Lau et al. 2009), and clumping of the gas (Morandi \& Cui 2014).
%Recent numerical studies \citep{Diemer.Kravtsov.2014, Oneil.et.l.2021} shows that the dark matter profile at large radii deviates from NFW or Einasto profiles widely adopted in literature as universal analytic profiles. 
%Observationally, the cluster outskirts are actively being studied in X-rays (refs) and optical wavelengths (refs) as well as via the SZ effect (refs).  Observations are very challenging in cluster outskirts,
%since the X-ray surface brightness drops below the back-
%ground level at large radii. Thanks to eROSITA observations we have ex-
%tended X-ray measurements of the intracluster medium ICM) profiles out to and beyond R 200 , where R 200 is the
%radius within which the mean total density is 200 times the
%critical density of the Universe.  

\section{Sample selection}
\label{sec:sample}

For the purpose of this study, we need a sample of clusters with known redshifts and reliable mass estimates. We further would like to suppress spatial variations of the X-ray sky backgrounds and foregrounds to get an accurate estimate of the excess emission associated with clusters, i.e. the stacked image should not be dominated by a few very bright objects. Simultaneously, we want the clusters to be spatially well-resolved and sufficiently bright. A combination of these requirements calls for a sample of massive nearby clusters, from which a few nearest objects are excluded.   
%The latter condition is a compromise between a number of photons (too distant clusters may not be properly described due to poor statistics) and the angular size of a cluster. 

The galaxy clusters from the CHEX-MATE catalogue (namely, Tier-1) \citep{chex-mate1, chex-mate2} appear to meet all our requirements.  The CHEX-MATE is an unbiased, signal-to-noise limited sample of 118 galaxy clusters detected by Planck via Sunyaev-Zel’dovich effect \citep{1972CoASP...4..173S}. It is composed of clusters at $0.05 < z < 0.2$ with masses $2\times 10^{14} M_{\odot} < M_{500c} < 9\times 10^{14} M_{\odot}$ from the PSZ2 catalogue \citep{2016A&A...594A..27P}. We restrict the sample to objects with a Galactic longitude $0^{\circ} < l < 180^{\circ}$ (available to us according to the data-sharing agreement) and a Galactic latitude $|b|> 15^{\circ}$ to avoid background variations due to proximity to the Galactic plane and the Cygnus X star formation region. For the same reason, we further exclude a few clusters located near the North Polar Spur. The final list of 38 clusters is presented in Table~\ref{tab:sample}. The sample median mass is equal to $ M_{500c} = 4.13 \cdot 10^{14} M_{\odot}$, the median redshift is $ z  = 0.113$, and the hydrogen column density is $ N_H = 2.17 \cdot 10^{20}$ cm$^{-2}$.

\section{Data and data processing}
Orbital observatory \textit{SRG} \citep[][]{2021A&A...656A.132S}, featuring two focusing X-ray telescopes, eROSITA \citep[][0.3--8.0 keV]{2021A&A...647A...1P} and ART-XC \citep[][4--30 keV]{2021A&A...650A..42P} was launched in July 2019 and started to perform the all-sky survey mission in December 2019. By now, four complete all-sky surveys have been completed, resulting in unprecedentedly deep X-ray maps of the whole sky being obtained. 

The eROSITA telescope consists of seven nearly identical independent modules, each having its own mirror system, a position-sensitive detector, and electronics \citep{2021A&A...647A...1P}, and  for the imaging analysis we use the data of all seven telescope modules (TMs).  
Here we use the data accumulated over four consecutive scans, with the total effective exposure amounting to {1225} seconds (i.e. $\approx8600$~s in equivalent exposure for one telescope module). 
%, while TMs 5 and 7 are excluded from the spectral analysis due to the possible impact of the optical light leak on their signal \citep[e.g.][]{2021A&A...647A...1P}. 
Initial reduction and processing of the data were performed using standard routines of the \texttt{eSASS} software \citep{2018SPIE10699E..5GB,2021A&A...647A...1P}, while the imaging and spectral analysis were carried out with the background models, vignetting, point spread function (PSF) and spectral response function calibrations built upon the standard ones via slight modifications motivated by results of calibration and performance verification observations \citep[e.g.][]{2021A&A...651A..41C,2023MNRAS.521.5536K}. 

For each galaxy cluster from Table~\ref{tab:sample}, we made an X-ray image in the 0.3--2.3~keV band of a region $10 R_{\rm 500c} \times 10R_{\rm 500c}$ centred on a cluster, where the values of $R_{\rm 500c}$ and $z$ taken from the original sample were used to set the angular size of the image. As outlined below, a few operations have been performed before adding the images to the stack.

\subsection{Removal of sources}
In the 0.3--2.3 keV band, the eROSITA detector's intrinsic background makes up about $\sim$33\% of the total count rate, while distant AGNs and the Galactic diffuse emission contribute $\sim$42\% and $\sim$25\%, respectively, in the areas with small $N_H$ (namely, for the sample median $N_H = 2.17 \cdot 10^{20}$ cm$^{-2}$). Of course, these numbers fluctuate from field to field and are also subject to temporal variations of the detector background.

To reduce fluctuations in surface brightness due to bright sources,  we removed compact sources with the flux exceeding $3\times 10^{-14}\, {\rm erg\,cm^{-2}\,s^{-1} }$ in the standard 0.5--2 keV energy band. 
The procedure for detection and modelling of the point sources is identical to the one implemented and described in \cite{2021A&A...651A..41C}. Fluxes are converted from the detection band 
%(0.4-2.3 keV) 
to the standard 0.5--2 keV energy band via a constant conversion factor, calculated for a typical CXB spectrum. Although some sources might have different spectra, e.g. stars, they are not numerous at the flux limit of the constructed source catalogues. The shape and normalization of the logN--logS distributions of the obtained catalogues are fully consistent with the earlier measurements and models. 

Similar procedure is applied to mildly extended sources \citep[with the characteristic radius $\lesssim3$ arcmin, i.e. $\lesssim$10 times the characteristic radius for the $\beta$-profile model of the core PSF, e.g. ][for the detailed description]{2021A&A...651A..41C}, which are detected and modelled with the $\beta$-profile \citep{1976A&A....49..137C} of the surface brightness together with the point sources at the first stage. The limiting flux for the mildly extended sources depends on their size and the background level for the locations of the clusters in our sample, and a rather conservative significance limit corresponding to the flux of $\approx10^{-13}$ erg s$^{-1}$ cm$^{-2}$ has been chosen to ensure its uniformity across the sample while keeping suppression of the shot noise from the individual sources at the required level.

After automatic detection, modelling and removal of the point and mildly extended sources, more extended sources, typically having sizes bigger than a few arcmin and irregular shapes, are visually identified on the smoothed residual images \citep[as exemplified by the supernova remnant candidates found blindly in the X-ray survey data, see ][]{2023MNRAS.521.5536K}. Those are not numerous and were masked manually after a few iterations with varying mask sizes.

\begin{figure*}
\centering
\includegraphics[angle=0,width=2\columnwidth]{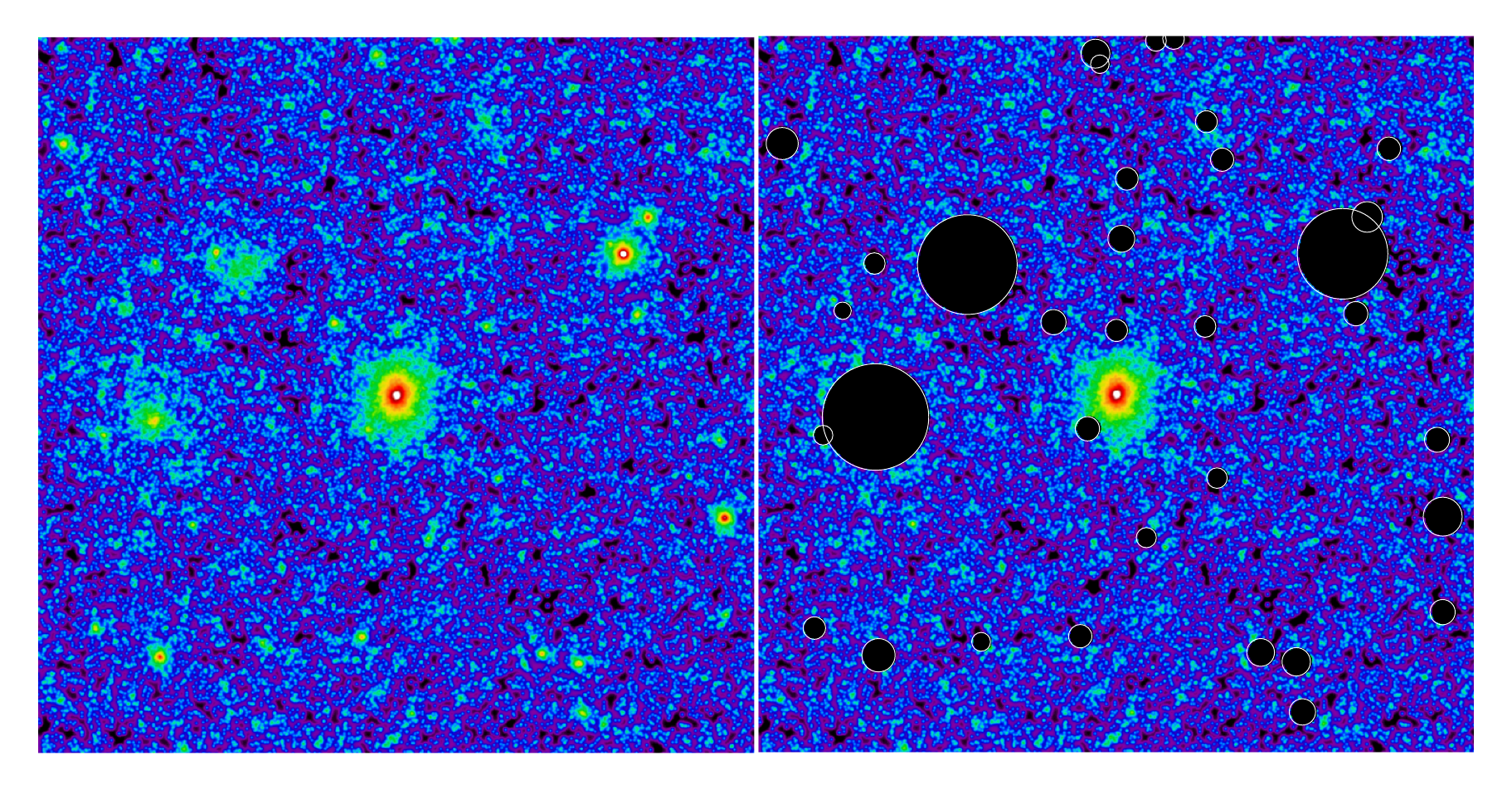}
\caption{Illustration of the sources removal procedure for the cluster G028.89+60.13.  The left panel shows the eROSITA image $5R_{\rm 500c} \times 5R_{\rm 500c}$
%in the 0.3-2.3 keV energy band 
(smoothed with the eROSITA PSF) prior to the removal of sources. 
%The mean (spatially flat) background has been subtracted from images to show the regions with excess emission more clearly. 
The right panel shows the same image with compact and extended sources masked. 
}
\label{fig:examples}
\end{figure*}

%Diffuse sources are also removed.
Fig.~\ref{fig:examples} illustrates the source removal procedure.  Fig.~\ref{fig:nomask} compares the stacked images with and without the removal of sources. The right panel of Fig.~\ref{fig:nomask} shows the major strength of the survey data when the intrinsic size of the telescope FoV is not imposing any constraints on the size of the studied region.

\subsection{Stray light} 
A small fraction of incoming photons can reach eROSITA detectors after one scattering by the telescope mirrors instead of  nominal two scatterings. These photons (usually called "stray light") produce spurious extended halos (up to $\sim 3$ degrees) around each X-ray source \citep[see, e.g. Fig.15 in][]{2021A&A...656A.132S}. To estimate the contribution of the stray light, we followed the approach described in Appendix~A of \cite{2023A&A...670A.156C}. Namely, every image has been convolved with a kernel that approximately characterizes the spatial distribution of stray light photons in the 0.4--2.3 keV band. The resulting convolved images are stacked similarly to cluster images and subtracted from them. This way a first-order correction for the stray light contribution is achieved.

\subsection{Stacking}

 To do actual stacking, we re-mapped every X-ray image as if the cluster was located at $D_{\rm ref}=343$ Mpc away from us and had $R_{\rm 500c} = 1\,{\rm Mpc}$ corresponding to the angular size of $\approx 10'$. The choice of these parameters is rather arbitrary and motivated largely by the desire to have a simple conversion of radii into the units of $R_{\rm 500c}$ or Mpc. The X-ray surface brightness was also adjusted to reflect the differences in distance, mass, redshift (see Appendix~\ref{app:epsc}) and the area distortions  due to the tangential (gnomonic) projection of individual images. Given that the redshift and mass dependencies are accounted for during the stacking procedure, the above reference distance $D_{\rm ref}$ is simply a coefficient that relates the physical and angular sizes in Euclidean space.  
 
 When doing stacking, we keep track of the photon counting noise and the estimated surface density of unresolved sources so that their contributions to the noise in the final images can be estimated. In parallel, the (properly scaled) exposure maps are also stacked in the same manner as the X-ray images.

%A pixel size is adjusted to make all images with the same resolution, i.e. the pixel size $\propto R_{500}$.  

%\begin{figure*}
%\centering
%\includegraphics[angle=0,width=1.9\columnwidth]%{figs/stacking_images_wwomask}
%\caption{Stacked exposure-corrected 0.3-2.3~keV image of 38 clusters with (right) and without (left) removal of resolved compact and extended sources. The images are $20\times R_{\rm 500c}$ on a side. The spatial uniformity of the background after removing sources is striking.}
%\label{fig:nomask}
%\end{figure*}

\section{Results}
\label{sec:res}

\begin{figure*}
\centering
\includegraphics[angle=0,width=2\columnwidth]{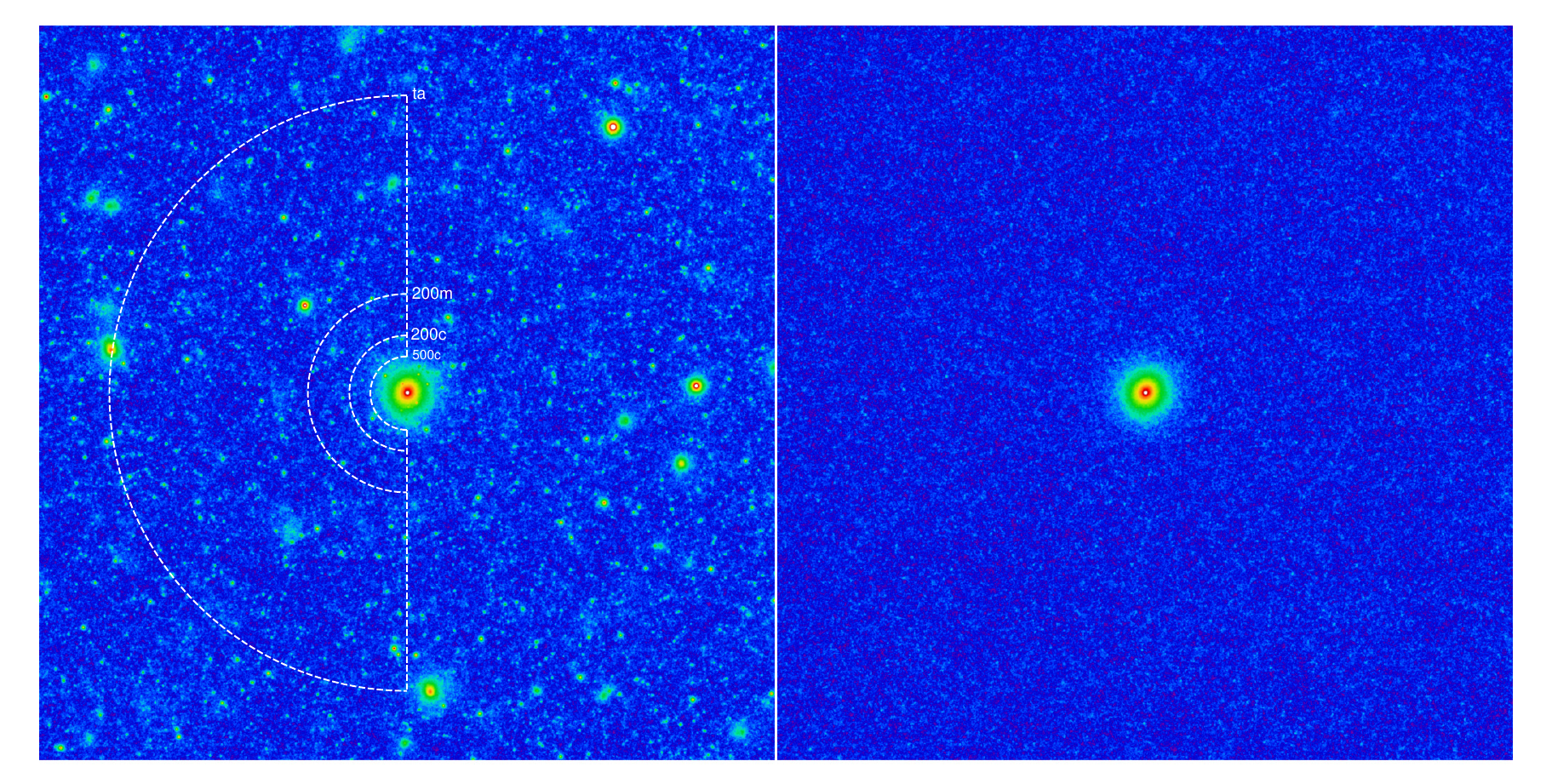}
\caption{Stacked exposure-corrected 0.3-2.3~keV image of 38 clusters with (right) and without (left) removal of resolved compact and extended sources. The images are $20\times R_{\rm 500c}$ on a side. The dashed semi-circles on the left image shows positions of $R_{\rm 500c}$, $R_{\rm 200c}$, $R_{\rm 200m}$, and the turn-around radius  $R_{\rm ta}$.  Note the striking spatial uniformity of the background after removing sources.}
\label{fig:nomask}
\end{figure*}

Fig.~\ref{fig:nomask} shows the final stacked image with a few characteristic radii indicated by white lines. Namely, we adopt the following relation between these radii  $R_{\rm ta}$:$R_{\rm 200m}$:$R_{\rm 200c}$:$R_{\rm 500c}$ = 8.1:2.7:1.6:1 according to  \cite{Nelson.et.al.2014, Diemer.Kravtsov.2014}. 
By design, $R_{\rm 500c}$ in the stacked image corresponds to 1~Mpc (or $10'$ at the adopted reference distance). Given that all images were originally made in equatorial orientation and then co-added, the final 2D image does not bear much information except for a demonstration that no prominent peculiar structures have been left in the image and the resulting image appears symmetric. Instead, most of the important information is contained in the radial surface brightness profile, which is related to the 3D emissivity profile. The latter profile is assumed to be spherically symmetric. However, 2D stacked images could be used to compare the expected and actual levels of noise by calculating the radial profile in several wedges as we do below.

\subsection{The X-ray surface brightness and gas density profiles}
To describe the observed surface brightness profile of the stacked image, one could use a modified $\beta$-model \citep[e.g.][]{2006ApJ...640..691V, Vikhlinin.et.al.2009} to parameterize the radial dependence of the emission measure, i.e. the product of the electron and proton number densities
\begin{equation}
    n_p n_e  = A \frac{(r/r_c)^{-\alpha}}{\left(1+\frac{r^2}{r_c^2}\right)^{3\beta - \alpha/2}} 
    \frac{1}{\left[1+ \left(\frac{r}{r_s}\right)^{\gamma} \right]^{\epsilon/\gamma}}.  
 \label{eq:beta}   
\end{equation}
In cosmological simulations, dark matter and gas density profiles in the outer cluster regions show varying logarithmic slope \citep[e.g.,][]{Diemer.Kravtsov.2014, Oneil.et.l.2021}. This behaviour is not fully captured by formula~(\ref{eq:beta}), so we further modify it by introducing an additional multiplier:
\begin{eqnarray}
    n_p n_e = & A \frac{(r/r_c)^{-\alpha}}{\left(1+\frac{r^2}{r_c^2}\right)^{3\beta - \alpha/2}} 
        \frac{1}{\left[1+ \left(\frac{r}{r_s}\right)^{\gamma} \right]^{\epsilon/\gamma}} \times   \frac{\left[1+\left(\frac{r}{r_{s2}}\right)^{\gamma}\right]^{\epsilon/\gamma}}{\left[1+\left(\frac{r}{r_{s2}}\right)^{\xi}\right]^{\lambda/\xi}} = \nonumber \\
        &A\times F(r,r_c,\alpha,\beta,r_s,\gamma,\epsilon,r_{s2},\lambda,\xi),  
 \label{eq:beta_mod}   
\end{eqnarray}
so that the resulting profile can change its slope near $r_s$ and  $r_{s2}$.
In total, the model has 10 free parameters. While the values of these parameters 
are  correlated, the Eq.~(\ref{eq:beta_mod}) provides us with sufficient flexibility to describe the radial behaviour of the emission measure found in simulations. For an X-ray emissivity that weakly depends on temperature (see below), the expected surface brightness profile is described by the line-of-sight integral of the above expression.

%needed to describe observed X-ray surface brightness profile up to several $R_{500}$.  
%We stress here that the formula above is flexible enough to reproduce the gas profile for simulated halos out to $\sim5R_{200m}$ \citep[see Fig. 5 in][]{Oneil.et.l.2021}. 

\begin{figure}
\centering
\includegraphics[angle=0,width=\columnwidth]{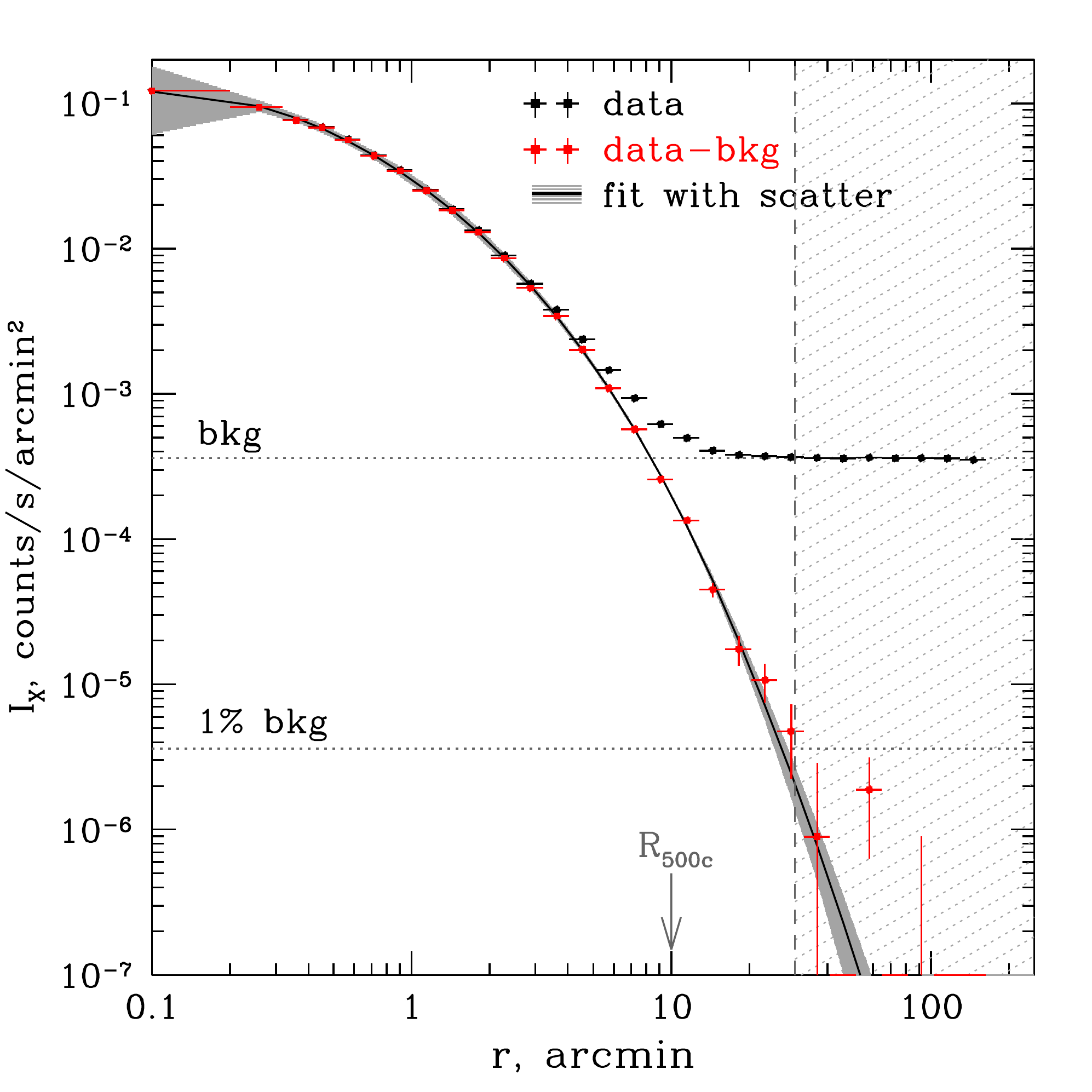}
\caption{The radial surface brightness profile of the stacked image in the 0.3-2.3 keV band (the black points). The red points show the same radial profile from which the best-fitting estimate of the constant background (upper horizontal dotted line) has been subtracted. The  black line shows the best-fitting model of the surface brightness profile (line-of-sight projection of Eq.~\ref{eq:beta_mod}). Grey shaded area shows estimated scatter in the best-fit model determined from dividing the stacked image into four 90-degree wedges and calculating the standard deviation of the best-fitting profiles in four wedges divided by 2 (see also Appendix~\ref{app:wedges}).  The vertical dashed line at $3\times R_{\rm 500c}$ shows the maximum radius where the excess X-ray emission has been detected. Beyond this radius, all values are considered as an extrapolation of our parametric model. Also, for comparison, the level of 1\% of the spatially uniform background is shown. }
\label{fig:sb}
\end{figure}

We extract the surface brightness profile from the stacked X-ray image (Fig.~\ref{fig:nomask}) and then fit it with the projected model given by Eq.~(\ref{eq:beta_mod}) plus a spatially constant background component. Fig.~\ref{fig:sb} shows the observed surface brightness profile together with the best-fitting model. The parameters of the model are given in Appendix~\ref{app:bfm}. The uncertainty in the constant background level is estimated to be of the level of $\approx$1\%. This value was obtained by dividing the stacked image into four 90-degree wedges and calculating the standard deviation between best-fitting constants in those wedges. %$R_{500}\equiv 10'$ is marked with the vertical dotted line.  The constant background level is shown with the horizontal dashed line. 

%The best surface brightness fit was converted to gas density by calculating an emissivity profile in \texttt{XSPEC}, taking into account the eROSITA instrumental response and assuming the absorption $N_H=10^{20}\,{\rm cm^{-2}}$, abundance $Z/Z_\odot=0.3$.  The gas density then $\rho_{gas} = \mu n_e m_p$, where $n_e$ is the electron density measured in X-rays, $m_p$ is the proton mass, and $\mu=1.148$ is the mean molecular weight per free electron. We compare the gas density obtained in this work with the profile presented in \cite{Oneil.et.l.2021} who derived a stacked gas density for simulated haloes between $10^{14}M_{\odot}$ and $10^{14.5}M_{\odot}$ at z=0 in TNG300. 

The best-fitting model to the surface brightness in units of ${\rm counts\,s^{-1}\,arcmin^{-2}}$ has been converted into the proton density\footnote{Throughout the paper we assume that hydrogen is fully ionized.} profile as follows
\begin{eqnarray}
n_p(r)=\left ( \frac{A F(r)}{\theta^3\epsilon_c D_{\rm ref}} \frac{n_p}{n_e} \right )^{1/2},
\label{eq:nhfromx}
\end{eqnarray}
where $\theta=2.9\times 10^{-4}$ corresponds to $1'$ and $n_p/n_e=0.83$. The value  $\epsilon_c=5\times 10^{-14} \,{\rm cm^5 counts~s^{-1}}$ used in the above expression  corresponds to a typical combination of cluster parameters in the sample (see Appendix~\ref{app:epsc} and Fig.~\ref{fig:epsilon}). The spread in the absorption column densities, mean temperatures, and redshifts among the clusters in the sample translates into variations of $\epsilon_c$ between $4$ and $5$. This translated into  $\sim$11\% uncertainty in the recovered typical gas density. A larger spread could be present when the entire radial range is considered (including the cool core regions or the very outskirts). For this pilot study, we keep the assumption of a constant $\epsilon_c$ and plan to do  a first-order correction using temperature and abundance profiles from numerical simulations in the subsequent work. A brief discussion on the expected level of uncertainties is given in Appendix~\ref{app:epsc}.

We also note here that in the outer regions of a cluster, several other effects can impact the X-ray signal. These include non-equilibrium ionization \citep[e.g.][]{2006PASJ...58..641Y}, photoionization \citep[e.g.][]{2001MNRAS.323...93C,2019MNRAS.482.4972K,2022MNRAS.515.3162S}, and resonant scattering of the line \citep[e.g.][]{1987SvAL...13....3G} and/or continuum photons. For the latter process, the difference in temperatures between the cluster core and outskirts implies that different ions are present in these two regions. Therefore,  the photons from the continuum will be scattered in the cluster's distant cluster outskirts. In this case, the expected signal is more than an order of magnitude below the surface brightness levels relevant to this study.

\subsection{Unresolved CXB sources}
Apart from pure photon-counting noise, additional variations in the derived surface brightness profiles could come from fluctuations of various backgrounds and foregrounds, and, also, from "peculiarities" of individual clusters. Some of these contributions can be straightforwardly included in the error budget. For instance, this is true for Poisson fluctuations in the number of unresolved sources that compose CXB. The surface density of unresolved sources was estimated and the expected noise level was corrected to include these fluctuations (single-halo approximation). Note that clustering of X-ray sources \citep[e.g.][]{1995ApJ...455L.109V} is not accounted for here, and may increase noise associated with unresolved X-ray sources.
To this end, the number counts of the \textit{Chandra} deep fields were used \citep{2017ApJS..228....2L}. 
%This procedure only accounts for the single-halo contribution to the noise. 
Given that we use the same threshold for compact sources in all fields, this type of noise scales with the solid angle of the studied region similarly to the photon counting noise, albeit its amplitude does not depend on the exposure time. Effectively, for this sample, the noise increases by a factor $\sim 1.9$ compared to the pure photon counting noise.

The comparison of wedge-to-wedge variations in the radial profile extracted from the final stacked image (four 90-degrees wedges were used) shows that the scatter is within a factor of less than 2 from expectations. Part of this extra noise comes from the remaining azimuthal asymmetries in the cluster core regions, where statistical errors are very small. In the most important radial range between $\sim 0.6$ and $3\times R_{\rm 500c}$, the scatter between wedges agrees well with expectations. We, therefore, concluded that other types of uncertainties, e.g. large-scale sky background variations, do not dominate the error budget.

%NH	3.1438042E20	2.0973749E20	8.82E19	8.43E20	2.559E20	46

%mean,stddev,min,max,median

\begin{figure}
\centering
\includegraphics[width=\columnwidth]{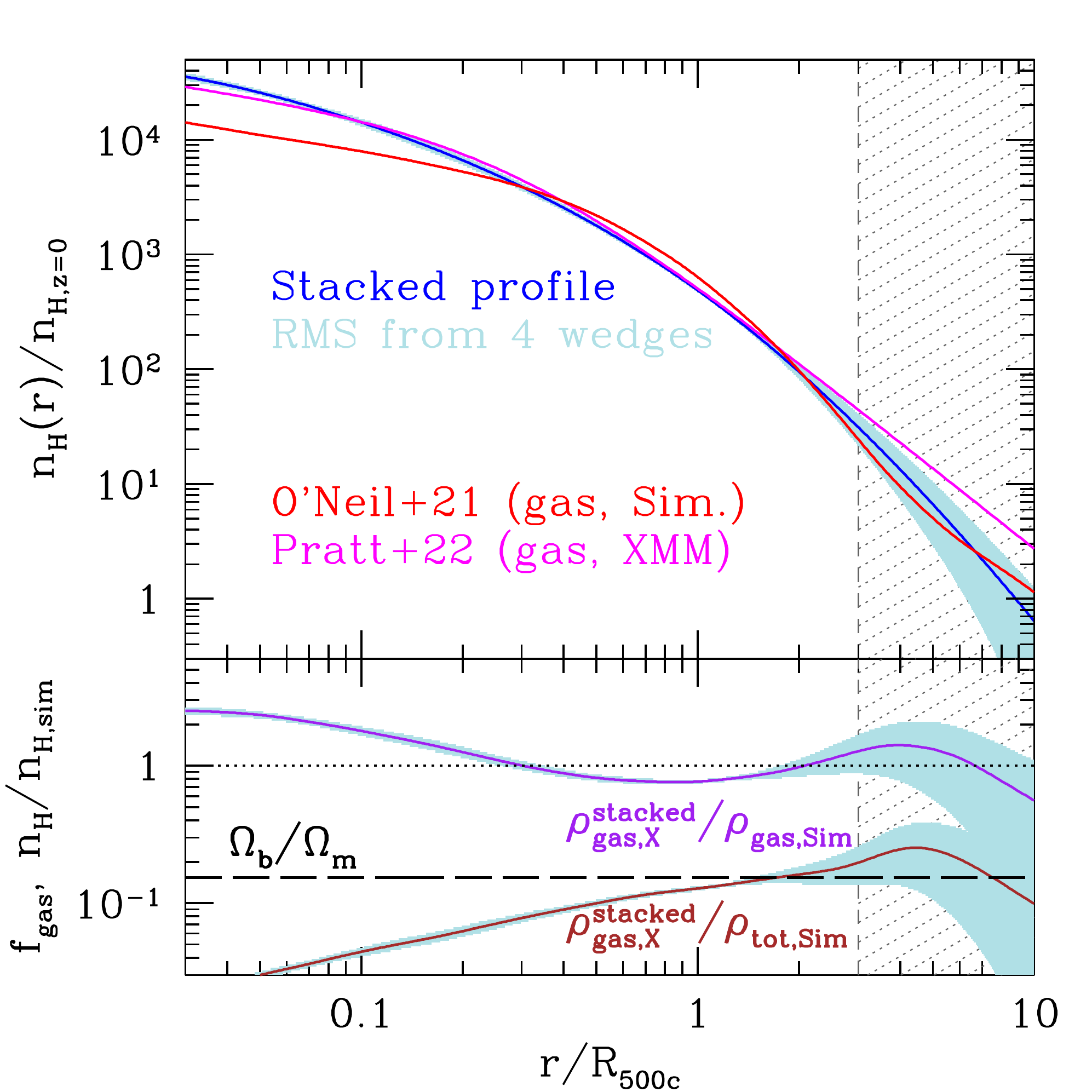}
\caption{Baryon density profile derived from the stacked X-ray image (blue). The light-blue curves show the density profiles in four independent $90^\circ$-wedges. For comparison, the magenta curve shows the density profile from \citealt{2022A&A...665A..24P, 2023A&A...669C...2P} (based on XMM-Newton observations) converted to the proton density. The density profile derived from  numerical simulations \citealt{Oneil.et.l.2021} is shown with the red line and demonstrates good agreement with the X-ray data. The ratio of the stacked gas density to the gas density from \citealt{Oneil.et.l.2021} is shown in the lower panel in purple. In the lower panel, we also plot the gas fraction (in brown) estimated as the ratio between the density profile found in this work to the total matter density from  \citealt{Oneil.et.l.2021}. The (local) gas fraction is close to the baryonic fraction $\Omega_b/\Omega_m$ beyond $\sim R_{\rm 500c}$. The vertical line shows the maximum radius where the excess X-ray emission above the background level is detected. The agreement of the measured density profile in terms of shape and normalization is good over the range of radii probed in X-rays. }
\label{fig:nh}
\end{figure}

\begin{figure}
\centering
\includegraphics[angle=0,trim=1cm 5cm 0cm 2cm,width=\columnwidth]{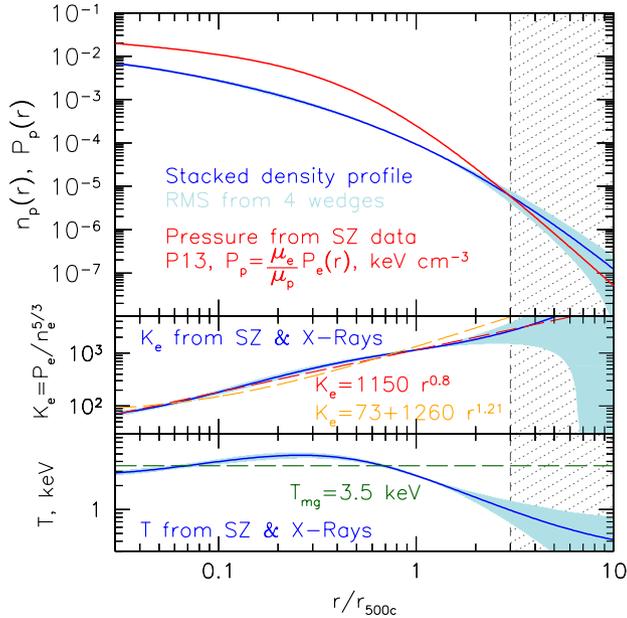}
 \caption{Gas temperature and entropy derived from the comparison of the pressure profiles based on the SZ data and the gas density profile obtained from the stacked X-ray image. The top panel shows a version of the universal pressure profile (the red line) from \citet{2013A&A...550A.131P}, that was converted to "proton" pressure as $P_{\rm p}=P_{\rm e}\times \mu_{\rm e}/\mu_{\rm p}$ in units of ${\rm keV\, cm^{-3}}$.  The proton density $n_{\rm p}$ derived from the stacked image (the same as in Fig.~\ref{fig:nh}) is shown with the blue line. The two lower panels show the entropy index $K_{\rm e}=P_{\rm e}/n_{\rm e}^{5/3}$ in units of ${\rm keV\,cm^{2}}$ and the gas temperature $T=P_{\rm p}/n_{\rm p}$ in keV. The red dashed line shows a single power law approximation of the entropy profile over the entire radial range probed here.  For comparison, the dashed orange line corresponds to the fit for the full ACCEPT sample \citep[Table 4 in][]{2009ApJS..182...12C}. While the functional forms are different, all curves agree within a factor of 2. In dashed horizontal line in the temperature plot shows the mass-weighted temperature $T_{\rm mg}$ from \citealt[][]{2006ApJ...640..691V} for $M_{\rm 500c}=2.85\times 10^{14}\,{\rm M_\odot}$.}
\label{fig:upp}
\end{figure}

\section{Discussion}
Given our choice\footnote{We remind a reader, that our choice of $R_{\rm 500c}=1\,{\rm Mpc} = 10 $ arcmin is motivated solely by the desire to have a simple conversion between radii in angular and physical units.} of $R_{\rm 500c}=1\,{\rm Mpc}$ as a fiducial value, 
the corresponding mass is $M_{\rm 500c}=500\rho_c\frac{4}{3}\pi R_{\rm 500c}^3\approx 2.85\times 10^{14}\,{\rm M_\odot}$. Below, we will use this mass to scale models for comparison with observational results.  

The radial gas density profile\footnote{Strictly speaking, the X-ray surface brightness profile.} that extends up to $\sim 3 \times R_{\rm 500c}$ is the most principal result of this study. For estimates, we assume that the uncertainties in the recovered density profile at a given radius can be characterized by the scatter between best-fitting model profiles at the same radius extracted in four 90-degree wedges (scaled down by a factor of 2). On top of this, comes the uncertainty in the  conversion coefficient $\epsilon_c$ discussed above. 

\subsection{Density}
The density profile (number density of protons normalized by the mean density $n_{p,z=0}=n_{{\rm H},z=0}=1.92\times 10^{-7}\,{\rm cm^{-3}}$ at $z=0$) corresponding to the best-fitting model is shown in Fig.~\ref{fig:nh} with the blue line (a set of best-fitting parameters is given in Appendix~\ref{app:bfm}). The light-blue shaded region shows the r.m.s. scatter between the best-fitting models in four $90^\circ$-wedges. This r.m.s. value was divided by 2, to show the expected amplitude of variations in the profile under the assumption that the variations of the surface brightness in the four wedges are statistically independent. The dashed vertical line marks the position of the last significant point in the surface brightness profile (see Fig.~\ref{fig:sb}). At this radius,  the r.m.s. is $\sim 0.3$ of the best-fitting model value.
%In the surface brightness profile, this radius corresponds to the last significant point. We consider the values beyond this radius as a pure extrapolation of our parametric model.

The gas density profiles around massive halos have been calculated in numerical simulations \citep[e.g.][]{2015ApJ...806...68L,Oneil.et.l.2021} and, also, derived from X-ray data \citep[e.g.][]{2006ApJ...640..691V,2008A&A...487..431C,2012A&A...541A..57E,2022A&A...665A..24P}.   For example, the red line in Fig.~\ref{fig:nh} shows the gas overdensity profile of the stacked halos  between $10^{14}$ and $10^{14.5}\,{\rm M_\odot}$ at $z=0$ in TNG300-1 from \citealt{Oneil.et.l.2021}, see their Fig.~5. The ratio of the two profiles is shown in the bottom panel with the purple line. Within the radial range $(\sim 0.2-3)\times R_{\rm 500c}$, the two profiles agree within $\pm30$\%.  The brown curve shows the ratio of the gas density (again normalized by $n_{{\rm p},z=0}$) from the X-ray data to the total matter overdensity from  \citealt{Oneil.et.l.2021}. As expected, this ratio (known as $f_{\rm gas}$) is close to the mean baryon fraction $\Omega_b/\Omega_m\approx 0.15$ beyond $R_{\rm 500c}$. 

In the same figure, we show the density profile model obtained by \citealt{2022A&A...665A..24P, 2023A&A...669C...2P} from the XMM-Newton observations of 118 clusters. 
%from the same sample ??? \ec{double check} (the magenta line). 
It shows an excellent agreement with the eROSITA-based profile in the range  $(0.03-1)\times R_{\rm 500c}$. At large radii ($R_{\rm 500c}<R<3 R_{\rm 500c}$), the models start to deviate from each other. In this regime, the extrapolation of   \citealt{2022A&A...665A..24P} model predicts higher density than derived from the eROSITA data.

\subsection{Temperature and entropy}
Measuring stacked spectra to determine the gas temperature beyond $R_{\rm 500c}$, is much more difficult than measuring the X-ray surface brightness in a broad energy band. However, one can use the ratio of pressure profiles derived from SZ data  to the density profile from X-ray data to measure the temperature. Various approximations of the pressure profiles (usually referred to as \textit{Universal Pressure Profile} or UPP) are available, starting from the work by \citealt{2010A&A...517A..92A}, which was based on X-ray data and uses a parametric model (generalized NFW) motivated by numerical simulations \citep{2007ApJ...668....1N}.  Subsequent studies \cite[e.g.][]{2013A&A...550A.131P,2013ApJ...768..177S,2021A&A...651A..73P,2023arXiv230409041M} used SZ data that can probe larger radii than typically mapped with X-rays. In the top panel of Fig.~\ref{fig:upp}, we plot the universal pressure profile from \citealt{2013A&A...550A.131P}
(the red line) scaled to our reference value of the mass $M_{\rm 500c}=2.85\times 10^{14}\,{\rm M_\odot}$. This pressure profile was converted from the electron pressure profile $P_{\rm e}$ to the "proton" pressure, i.e. $P_{\rm p}=n_{\rm p} kT=\frac{\mu_{\rm e}}{\mu_{\rm p}}P_{\rm e}$, where $\mu_{\rm e}$ and $\mu_{\rm p}$ are the mean atomic weights per electron and proton, respectively. For comparison, we plot the density profile $n_{\rm p}(r)$ from Fig.~\ref{fig:nh} with the blue line. The ratio of these two curves is the effective temperature that is shown in the bottom panel. For comparison, the horizontal dashed line shows the expected value of temperature from the scaling relation of \citealt[][]{2006ApJ...640..691V} for $M_{\rm 500c}=2.85\times 10^{14}\,{\rm M_\odot}$. This value corresponds to the mass-weighted temperature $T_{\rm mg}$ that on average is some 20\% lower than the peak temperature in the cluster radial profile.  In the derived temperature profile, the broad peak of $T\sim 4.8\,{\rm keV}$ is near $\sim 0.25 R_{\rm 500c}$.  Within $R_{\rm 500c}$ the temperature profile agrees with typical trends found in observations of nearby clusters and simulations \citep[e.g.][]{1998ApJ...503...77M,2002ApJ...567..163D,2002ApJ...579..571L}. 
At $3\times R_{\rm 500c}$, the temperature drops to $\sim 1\,{\rm keV}$, i.e. a factor of $\sim 5$ lower than the peak value.
Other variants of the UPP yield qualitatively similar results, although the scatter is substantial. In particular, the wiggles in the pressure profile found by \citealt{2022MNRAS.514.1645A} are just beyond $3\times R_{\rm 500c}$.  This scatter will be discussed in more detail in the subsequent work. 

The gas entropy profile is yet another important quantity that characterizes the history of gas heating via shocks and other dissipative processes  \citep[e.g.][]{2001ApJ...546...63T, 2005MNRAS.364..909V,2009ApJS..182...12C,2010A&A...511A..85P,2012MNRAS.427L..45W,2019A&A...621A..41G}. 
The same approach of combining the X-ray and SZ data can be used to estimate the entropy of the gas. We follow the standard convention and show the value of $K_{\rm e}=P_{\rm e}/n_{\rm e}^{5/3}$ in units of ${\rm keV\,cm^2}$ in the middle panel of Fig.~\ref{fig:upp}. The radial dependence of entropy across the radial range shown in Fig.~\ref{fig:upp} can be approximated with a power law  $K_{\rm e}=1150 \left ({r/R_{\rm 500c}} \right )^{0.8}\,{\rm keV\,cm^2}$ (the dashed red line). While this dependence is shallower than expected in non-radiative models \citep{2001ApJ...546...63T, 2005MNRAS.364..909V}, deviations from the steeper variants of the entropy profile approximation (outside the innermost region) are within a factor of $\sim 2$.  For example, the dashed orange line corresponds to $K_{\rm e}=73+1260 \left ({r/R_{\rm 500c}} \right )^{1.21}\,{\rm keV\,cm^2}$, which is based on the full ACCEPT sample \cite[see Table 4 in][]{2009ApJS..182...12C}. 

%While 
%. This dependence is consistent with the radial entropy floor 
% A by-eye fit $K_{\rm e}=10^3\, \left ({r/R_{\rm 500c}} \right )^{0.75}\,{\rm keV\,cm^2}$ shown with the dashed line is shallower than expected in non-radiative models \citep{2001ApJ...546...63T, 2005MNRAS.364..909V}.  We defer the detailed discussion of the radial profile for future study but note the smooth behaviour of the radial profile at large radii.

\begin{comment}

Voit fits to sim $K\propto r^{1.1}$. From Pratt+2010 $K=1.42*966*(r/r_{500})^{1.1}$ - this is for sims.
But real profiles

\begin{eqnarray}
T_X=5\,{\rm keV} \left ( \frac{M_{500c}}{3\times 10^{14}h^{-1}\,M_\odot}\right ) ^{0.65} E(z)^{0.65}
\end{eqnarray}
\citep{Vikhlinin.et.al.2009}

Temperature profiles from Maxim (obs) and simulations

In Loken+2002
\begin{eqnarray}
T(r)=T_X\left ( 1+\frac{r}{a_X}\right )^{-1.6},
\end{eqnarray}
where $a_X=r_{vir}/1.5$ and $r_{vir}$ corresponds to $\Delta_c=101$.
\end{comment}

\subsection{Clumpiness}
Gas clumpiness\footnote{In more general terms - any deviations from a spherically symmetric gas distribution that can not be unambiguously corrected in projected images.} is one of the issues that complicate the conversion of the X-ray surface brightness to the gas density. In the simplest case, when the emissivity does not depend on temperature, the X-ray surface brightness depends on $\langle n^2 \rangle$ that is larger than $\langle n \rangle^2$, which makes the density estimates $\propto \langle n^2 \rangle ^{1/2}$ biased high. Some  level of clumpiness is inevitably present both in the simulations and real objects and can affect all thermodynamic properties, although the magnitude of the effect is still a question of debate \citep[e.g.][]{2011Sci...331.1576S,2013MNRAS.428.3274Z,2015MNRAS.446.2629E}. 
In particular, numerical simulations suggest that the Probability Density Distribution (PDF) of the gas density in radial shells can be well described by a lognormal distribution, once a small number of high-density clumps is removed \citep[e.g.][]{2007ApJ...659..257K,2013MNRAS.428.3274Z,2013MNRAS.431..954K}, i.e.
\begin{eqnarray}
{\rm PDF}(n)=\frac{1}{\sqrt{2\pi}\sigma} e^{\frac{-(\ln n)^2}{2\sigma^2}}\frac{1}{n},
\end{eqnarray}
where $n$ is the gas density divided by the median density value. In this case, the clumpiness factor 
\begin{eqnarray}
C=\frac{\langle n^2\rangle}{\langle n \rangle^2}=e^{\sigma^2} > 1.
\end{eqnarray}
The bias in the value of the gas density derived from X-ray flux is equal to $C^{1/2}=e^{\sigma^2/2}$. The known uncertainties shown by shaded regions in the bottom panel of Fig.~\ref{fig:nh} allow for a factor of 2 larger densities derived from X-ray data compared to that found in numerical simulations. This corresponds to the value of  $\sigma=\sqrt{2\ln 2} \approx 1.2$. Typical values of $\sigma$ found in numerical simulations at $(2-2.5)\times R_{500c}$ are in the range of 0.4-0.6 (see Fig.4 in \citealt{2013MNRAS.428.3274Z} or Fig.4 in \citealt{2013MNRAS.431..954K}). Therefore, the derived value can be considered as a confirmation that much more extreme levels of X-ray-emitting gas clumpiness are not required by the data. The curve  $\displaystyle \rho^{\rm stacked}_{\rm gas,X}/\rho_{\rm gas,Sim}$ shown in Fig.~\ref{fig:nh} remains below $\sim2$ at all radii, implying that a rough upper limits  of $\sigma\lesssim 1.2$ applies to the entire radial range probed. We note here that the calculated ratio $ \displaystyle \rho^{\rm stacked}_{\rm gas,X}/\rho_{\rm gas,Sim}$  is better suited to demonstrate overall agreement between observations and simulations (and set crude upper limits on the gas clumpiness), than to derive the actual level of clumpiness at a given radius.

A more elaborate version of the same estimate could include a possible correlation of the density and temperature variations \cite[e.g.][]{2013MNRAS.431..954K}. However, this is important only if there are strong variations of the emissivity with temperature. Given the effective (i.e. calculated taking into account eROSITA's response function) emissivity curves shown in the Appendix \ref{app:epsc}, it is clear that the departures towards higher temperatures are unlikely to affect the above conclusions. On the other hand, for deviations towards a much smaller temperature, say below 0.1~keV, the effective emissivity drops dramatically, effectively reducing the contribution of very cold clumps to the chosen X-ray band. Therefore, the above conclusion that the X-ray-emitting gas is not extremely clumpy stays unchanged. 
%\ec{EC: predlagayu sleduyushee predlozhenie ubrat'.} A much larger sample (or much deeper observations of individual objects) would be needed to bring the uncertainties in the derived density to the level of $\sigma\sim 0.6$ to challenge numerical simulations. 
A much larger sample combined with a matching sample of simulated clusters would be needed to (i) bring down the uncertainties at large radii, (ii) verify the consistency of directly observed quantities, e.g. surface brightness in observations and simulations, and (iii) verify that the level of clumpiness found in simulations can be robustly recovered over the range or radii.

%As is evident from Fig.~\ref{fig:epsilon2}, the assumption of an approximately constant emissivity in the 0.3--2.3~keV band is valid for gas temperatures above  $\sim 0.4-0.5$~keV, the condition that is likely violated outside $3\times R_{\rm 500c}$, but seems to be good for the bulk of the gas inside this radius, unless the temperature fluctuates by a factor much larger than 2. Furthermore, for the temperature and entropy profiles derived from the comparison of X-ray and SZ data, the covariance of the density and temperature fluctuations is important to predict the sign and the magnitude of the bias induced by fluctuations \citep{2013MNRAS.431..954K}. Here we simply mention that neither density  nor temperature profiles derived in this work require strong clumpiness of the gas. Indeed, the gas fraction at large radii is comfortably consistent with the universal baryon fraction\footnote{Adding the expected contribution due to stars \citep[e.g.][]{2018AstL...44....8K} will not affect this conclusion.} (Fig.~\ref{fig:nh}), while the temperature and entropy do not show any prominent departures from a power-law-like behavior up to $3\times R_{\rm 500c}$. A conservative statement is therefore that at the level of accuracy achieved here, no need for significant gas clumpiness is needed, although a combination of several effects might conspire to make the profiles smooth.  

\subsection{Can we go beyond $3\times R_{\rm 500c}$?}
Fig.~\ref{fig:nh} and \ref{fig:upp} suggest that X-rays remain a good proxy of the gas at least up to $\sim 3\times R_{\rm 500c}$. At some larger distances from the cluster, the X-ray signal will plausibly be dominated by a highly inhomogeneous combination of individual halos (e.g. groups of galaxies) and the warm-hot intergalactic medium emission. The signal there will also reflect a complicated structure of the accretion/merger shocks \citep[][]{2009A&A...504...33V,2021MNRAS.508.2071A,2021MNRAS.506..839Z} and/or a non-equilibrium ionization \citep[e.g.][]{2006PASJ...58..641Y}. At even larger distances, e.g. comparable to the cluster turn-around radius, the excess X-ray signal may simply reflect the overall overdensity of all "objects" compared to volume averaged quantities for a representative chunk of the Universe.  In other words, the X-ray signal might switch from the quadratic dependence of the baryon density $\propto \rho_{\rm gas}^2$ in the inner part to a more linear dependence   $\propto \rho_{\rm gas}$ in the far outskirts. On top of this, come the changes in gas emissivity.

It is nevertheless interesting to ask, what are the prospects of detecting the gas beyond  $3\times R_{\rm 500c}$ if the density follows predictions of numerical simulations, and the emissivity of the gas does not change dramatically, i.e. the X-ray signal is set by the line-of-sight integral $\epsilon_c \rho_{\rm gas}^2$, where $\epsilon_c\approx {\rm const}$ is the same as below  $3\times R_{\rm 500c}$. At $3\times R_{\rm 500c}$ the local overdensity is $\sim 25-30$ and it declines approximately as $R^{-3}$. Assuming that the noise scales as $\propto N^{-1/2}A^{-1/2}$, where $N$ is the size of the sample, and 
$A=2\pi \frac{\delta R}{R}R^2$ is the area of annulus with the radius $R$ and a constant  relative thickness $\frac{\delta R}{R}$. Since the surface brightness declines as $\rho^2 R\propto R^{-5}$, the size of the sample should increase as $N\propto R^8$.  Therefore, in order to go from  $\sim 3\times R_{\rm 500c}$ to $\sim 4\times R_{\rm 500c}$, the size of the sample has to be increased by an order of magnitude. While still feasible, it is clear that at these distances the detection of the proper X-ray signal will be very difficult against various systematic effects.  Nevertheless, doing a larger sample makes sense in order to reduce the noise near $3\times R_{\rm 500c}$. This will be a subject of the subsequent publication. The plan is to supplement the above analysis with stacking that uses  $R_{\rm 200m}$  instead of $R_{\rm 500c}$, which might be more appropriate for the outskirts.

%\ec{EC: A variant of Olbers' paradox for clusters: SHALL WE DISCUSS IT HERE? It was surely discussed somewhere else. $\int \frac{dN}{dz dM} \left( \frac{R_{\rm 500c}}{D_a}\right)^2 dz dM$ or something slightly more complicated with account for $\epsilon_c$, etc. } 

As a caveat, we mention that the full calibration of the stray light profiles is yet to be completed and, therefore, some adjustments in the outer radial bins are possible in the future. Here, the main conclusion is that with eROSITA it is possible to reach levels $\lesssim 1$\% of the sky background in the 0.3--2.3~keV that are needed to place constraints on the X-ray emission near $3\times R_{\rm 500c}$. With this conclusion in mind, it is possible to consider larger samples and a fully "forward" approach to the fitting procedure for a sample of cluster profiles rather than the stacked image approach adopted here.

%\ec{EC: TODO: internal check of the (core-)PSF role for quoted profiles. We are not interested in the cores, but need to check the magnitude of the effect in the outskirts}

\section{Conclusions}
%\nl{not done yet}
Exploiting the "unlimited" field-of-view of SRG/eROSITA in the all-sky survey, we stacked X-ray images of 38 nearby galaxy clusters drawn from the CHEX-MATE sample.
%after removing bright compact and extended sources. 
The images cover the radial range up to the turn-around radii of these clusters. The excess X-ray emission on top of the sky background is detected up to $\sim 3\times R_{\rm 500c}$. The recovered gas density profile agrees well with the results of cosmological simulations in the standard $\Lambda$CDM model and does not show evidence of strong gas clumpiness (for the gas with temperatures in the range relevant to the X-ray band). Combining the density profile with the universal pressure profile from SZ data yields the temperature and entropy profiles that do not feature any strong change in their radial trends up to the maximum radius probed.

\section*{Acknowledgements}
We are grateful to the referee for careful reading of the manuscript and valuable suggestions and comments.

This work is based on observations with the eROSITA telescope onboard \textit{SRG} space observatory. The \textit{SRG} observatory was built by Roskosmos in the interests of the Russian Academy of Sciences represented by its Space Research Institute (IKI) in the framework of the Russian Federal Space Program, with the participation of the Deutsches Zentrum für Luft- und Raumfahrt (DLR). The eROSITA X-ray telescope was built by a consortium of German Institutes led by MPE, and supported by DLR. The SRG spacecraft was designed, built, launched, and operated by the Lavochkin Association and its subcontractors. The science data are downlinked via the Deep Space Network Antennae in Bear Lakes, Ussurijsk, and Baikonur, funded by Roskosmos. The eROSITA data used in this work were converted to calibrated event lists using the eSASS software system developed by the German eROSITA Consortium and analysed using proprietary data reduction software developed by the Russian eROSITA Consortium. AAS was partly supported by the project number 0033-2019-0005 of the Russian Ministry of Science and Higher Education. IK acknowledges support by the COMPLEX project from the European Research Council (ERC) under the European Union’s Horizon 2020 research and innovation program grant agreement ERC-2019-AdG 882679.
%%%%%%%%%%%%%%%%%%%%%%%%%%%%%%%%%%%%%%%%%%%%%%%%%%
\section*{Data Availability}
X-ray data analysed in this article were used with the permission of the Russian SRG/eROSITA consortium. The data will become publicly available as a part of the corresponding SRG/eROSITA data release along with the appropriate calibration information. 
All other data are publicly available and can be accessed at the corresponding public archive servers.

%%%%%%%%%%%%%%%%%%%% REFERENCES %%%%%%%%%%%%%%%%%%

% The best way to enter references is to use BibTeX:

\bibliographystyle{mnras}
\nocite{2022A&A...665A..24P}
\bibliography{example} % if your bibtex file is called example.bib

% Alternatively you could enter them by hand, like this:
% This method is tedious and prone to error if you have lots of references
%\begin{thebibliography}{99}
%\bibitem[\protect\citeauthoryear{Author}{2012}]{Author2012}
%Author A.~N., 2013, Journal of Improbable Astronomy, 1, 1
%\bibitem[\protect\citeauthoryear{Others}{2013}]{Others2013}
%Others S., 2012, Journal of Interesting Stuff, 17, 198
%\end{thebibliography}

%%%%%%%%%%%%%%%%%%%%%%%%%%%%%%%%%%%%%%%%%%%%%%%%%%

%%%%%%%%%%%%%%%%% APPENDICES %%%%%%%%%%%%%%%%%%%%%

\appendix

%\clearpage

\section{List of clusters}

Table~\ref{tab:sample} lists 38 clusters drawn from the  CHEX-MATE sample  \citep{chex-mate1} used for  stacking. Selection criteria are discussed in \S\ref{sec:sample}.

\begin{table}
\centering
\renewcommand{\arraystretch}{1.3}
\caption{Subsample of the CHEX-MATE (Tier-1) clusters  with Galactic longitudes in the range $0<l<180^\circ$ and Galactic latitudes $|b|>15^{\circ}$. A few more objects projecting onto the North Polar Spur region were omitted from the sample (see \S\ref{sec:sample}). Values of $M_{\rm 500c}$ are taken from the PSZ2 catalogue. }
\begin{tabular}{|l|cc|cc|cc|cc|}
\hline
     name        &   $M_{\rm 500c}, 10^{14} M_{\odot}$       &       $R_{\rm 500c}$, kpc & z        \\ \hline
%G006.49+50.56 & 7.05 &  1336.8 &    0.0766 \\
%G021.10+33.24 & 7.79 &  1364.8 &    0.1514 \\
%G028.63+50.15  &  3.17  &  1021.6  &  0.0916   \\
G028.89+60.13 &	4.47 &	1133.8 &	0.1530 \\
G031.93+78.71  &  2.72  &   973.8  &  0.0724   \\
G033.81+77.18 &	4.46 &	1150.2 &	0.0622 \\
G040.03+74.95  &  2.34  &   927.8  &  0.0612   \\
G040.58+77.12  &  2.57  &   955.2  &  0.0748   \\
G041.45+29.10  &  5.41  &  1203.1  &  0.1780   \\
G042.81+56.61  &  4.22  &  1127.4  &  0.0723   \\
G044.20+48.66  &  8.77  &  1434.7  &  0.0894   \\
G046.88+56.48  &  5.10  &  1192.5  &  0.1145   \\
G048.10+57.16  &  3.54  &  1062.3  &  0.0777   \\
G049.22+30.87 &	5.90 &	1241.3 &	0.1644 \\
G049.32+44.37  &  3.76  &  1080.4  &  0.0972   \\
G050.40+31.17  &  4.22  &  1110.2  &  0.1640   \\
G053.53+59.52  &  5.21  &  1201.3  &  0.1130   \\
G056.77+36.32  &  4.34  &  1133.7  &  0.0953   \\
G057.61+34.93  &  3.70  &  1077.6  &  0.0802   \\
G057.78+52.32  &  2.32  &   924.6  &  0.0654   \\
G057.92+27.64  &  2.66  &   966.1  &  0.0757   \\
G062.46-21.35 &	4.11 &	1100.9 &	0.1615 \\
G066.68+68.44 &	3.80 &	1072.2 &	0.1630 \\ 
G067.17+67.46 &	7.14 &	1321.2 &	0.1712 \\
G067.52+34.75  &  4.49  &  1131.1  &  0.1754   \\
%G068.22+15.18  &  2.14  &   901.3  &  0.0567   \\
G071.63+29.78  &  4.13  &  1103.7  &  0.1565   \\
%G075.71+13.51  &  8.74  &  1440.8  &  0.0557   \\
G077.90-26.63 &	4.99 &	1177.4 &	0.1470 \\
G080.16+57.65  &  2.51  &   945.4  &  0.0878   \\
%G080.37+14.64  &  3.13  &  1016.2  &  0.0980   \\
G080.41-33.24  &  3.77  &  1079.6 &  0.1072   \\
G083.86+85.09  &  4.74  &  1150.2  &  0.1832   \\
G085.98+26.69  &  4.17  &  1102.9  &  0.1790   \\
G094.69+26.36  &  3.08  &   999.8  &  0.1623   \\
G098.44+56.59  &  2.83  &   977.1  &  0.1318   \\
G099.48+55.60  &  2.75  &   972.2  &  0.1051   \\
G105.55+77.21  &  2.20  &   907.4  &  0.0720   \\
G111.75+70.37  &  4.34  &  1116.9  &  0.1830   \\
G113.29-29.69  &  3.57  &  1060.1  &  0.1073   \\
G114.79-33.71  &  3.79  &  1083.8  &  0.0940   \\
G124.20-36.48  &  7.25  &  1322.0  &  0.1971   \\ 
G149.39-36.84  &  5.35  &  1200.3  &  0.1700   \\
G172.74+65.30  &  2.39  &   931.7  &  0.0794   \\
%G179.09+60.12 &	3.84 &	1080.8 &	0.1372 \\ 
\hline 
\end{tabular}
\label{tab:sample}
\end{table}

\section{Re-scaling images for stacking}
\label{app:epsc}
 %\ec{TO BE MODIFIED ACCORDING TO $E(z)$ scaling, as in imgstackfast36}
%\ec{NOT YET READY}
The observed energy flux from a volume element $dV$ of a cluster in the energy range between $E_1$ and $E_2$ is
\begin{equation}
    \int^{E_2}_{E_1} F(E) dE=\int_{E_1^{'}}^{E_2^{'}}\frac{\Lambda(E^{'},T)dE^{'}}{4\pi D^2_a (1+z)^4} n_en_p dV,
\end{equation}
where $\Lambda(E^{'},T)$ is the gas emissivity of the cluster at the rest-frame energy $E'$, $n_e$ and $n_p$ are the electron and proton densities, respectively, $D_a$ is the angular diameter distance, and $z$ is the redshift.  
Therefore, 
\begin{equation}
F(E) = \frac{\Lambda(E^{'},T)}{4\pi D^2_a (1+z)^4} \frac{dE^{'}}{dE} n_en_p dV = \frac{\Lambda(E^{'},T)}{4\pi D^2_a (1+z)^3}n_en_p dV,
\end{equation}
since $E^{'}=E(1+z)$.
In terms of the surface brightness (in counts per second per unit solid angle) recorded by a telescope with effective area $A(E)$
\begin{eqnarray}
I_{[E_1,E_2]} = \int ^{E_2}_{E_1} \int A(E) \frac{F(E)}{E} n_e n_p dl dE= \nonumber \\ \frac{1}{4\pi (1+z)^3} \int^{E_2}_{E_1} A(E) \frac{\Lambda(E^{'},T)}{E} dE \int n_e n_p dl = \nonumber \\ \epsilon_c(T,z,E_1,E_2) \int n_e n_p dl,
\end{eqnarray}
where $l$ is the distance along the line of sight and $\epsilon_c(T,z,E_1,E_2)$ for eROSITA is given in \cite[][see their appendix C]{2021A&A...651A..41C}\footnote{We note here, that the quantity $\frac{1}{ (1+z)} \int^{E_2}_{E_1} A(E) \frac{\Lambda(E^{'},T)}{E} dE$ is plotted there.}.

\begin{figure}
%\centering
\includegraphics[angle=0,trim=1cm 5cm 0cm 2cm,width=0.99\columnwidth]{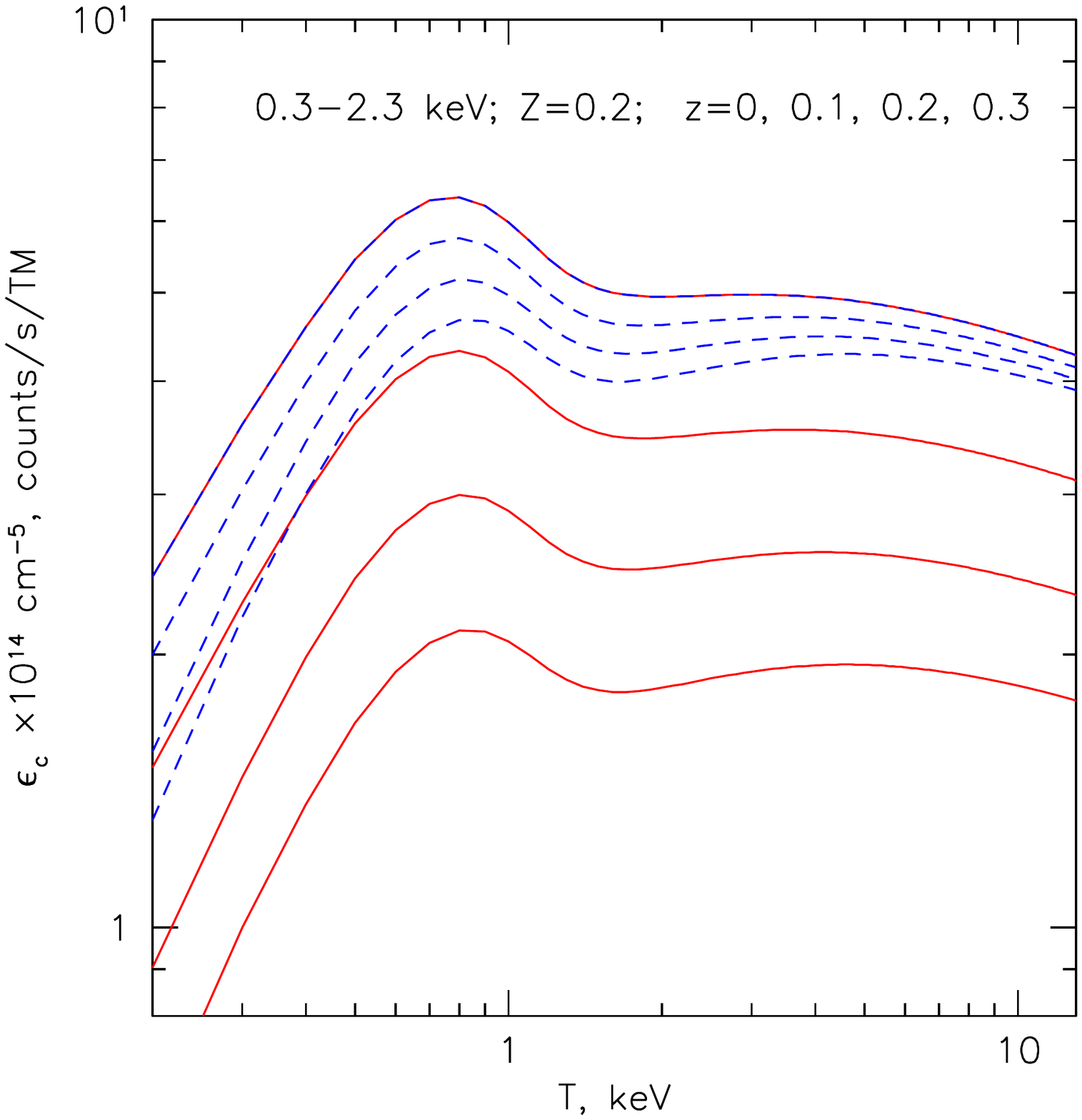}
\caption{The coefficient $\epsilon_c(T,z,E_1,E_2)$  that relates the line-of-sight-integrated emission measure and the observed surface brightness for a one (out of seven) eROSITA telescope module (TM) in the 0.3-2.3~keV band for gas metallicity  $Z/Z_\odot=0.2$, Galactic hydrogen column density $N_H=10^{20}\,{\rm cm^{-2}}$, and several values of the redshift $z$ (red curves, the top one corresponds to $z=0$, while the curve in the bottom is for $z=0.3$). The blue dashed lines show the same coefficient multiplied by $(1+z)^3$ to partly correct for the effects of cosmological dimming (including time dilation) of the X-ray signal. This  factor, $(1+z)^3$, is applied to clusters in the process of stacking. For our selection criteria, namely $z<0.2$, the variations of the corrected value of $\epsilon_c$ (dashed curves) with $z$ translates into $\sim 8$\% uncertainty in the derived density. Other factors included in Eq.~\ref{eq:corr} correct for the cluster's physical size and the redshift-dependent critical density.}
\label{fig:epsilon}
\end{figure}

\begin{figure}
%\centering
\includegraphics[angle=0,trim=1cm 5cm 0cm 2cm,width=0.99\columnwidth]{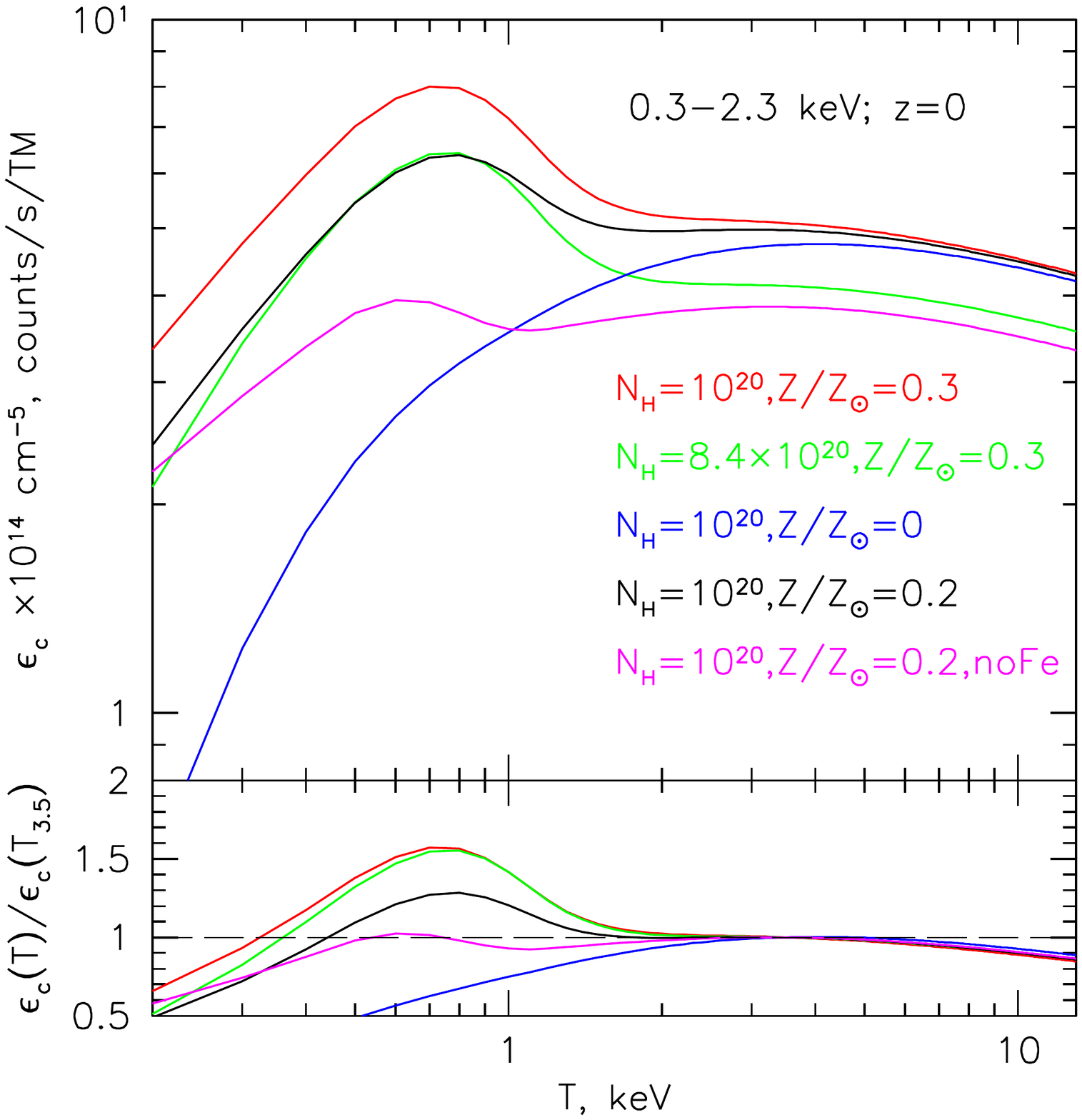}
\caption{Impact of variations of the metallicity $Z/Z_\odot$ and the Galactic hydrogen column density $N_H$ on the value of $\epsilon_c(T,z,E_1,E_2)$ for the 0.3-2.3~keV band. The recovered value of gas density scales as $1/\sqrt{\epsilon_c}$. The top panel shows the temperature dependence of $\epsilon_c$ for given values of $N_H$ and $Z/Z_\odot$. The bottom panel shows the same curves normalized by the value of $\epsilon_c$ at the reference temperature $T=3.5\,{\rm keV}$. Rather extreme values of $N_H$ and $Z/Z_\odot$ are used here to illustrate their impact. Increasing $N_H$ (compare the red and green lines) shifts the curves down, but does not affect the temperature dependence. Increasing the metallicity (compare the red and blue curves) affects the emissivity below $kT\sim 2\, {\rm keV}$. A characteristic "bump" appears at temperatures $\sim 0.7-0.8\;{\rm keV}$ largely associated with the Fe~L complex. The amplitude of this bump can be reduced by excluding the photons in the interval of energies 0.8-1.1 keV (see the magenta curve) at the cost of the reduction in the total signal. The black curve ($N_H=10^{20}\,{\rm cm^{-2}}$ and $Z/Z_\odot=0.2$) can be considered as a "fiducial" case for clusters' outskirts and is used in our analysis.}
\label{fig:epsilon2}
\end{figure}

When stacking cluster images with different masses and at different redshifts, the aim is to rescale the angular size and the observed surface brightness so that if these clusters obey the most simple form of scaling relations they will appear similar to each other. In particular, one can relate the mass to some characteristic radius and adjust the angular size and density accordingly. While for clusters' outskirts scaling with radii at fixed matter overdensity is more appropriate, e.g. $R_{200m}$ \cite[see, e.g.][]{Oneil.et.l.2021}, we used here the radii at fixed overdensity relative to the critical density of the Universe at a given redshift, namely $R_{\rm 500c}$. Since the mass and redshift ranges used here are small $0.05<z<0.2$, the impact of choosing a different scaling radius is expected to be small. Accordingly, the observed surface brightness $I_{obs}$ for every cluster was renormalized assuming that the typical gas density at $R_{\rm 500c}$ is $\propto E(z)^2$ and the physical size is $\propto R_{\rm 500c}$. Thus,
\begin{eqnarray}
I=I_{obs}\dfrac{r_{ref}}{R_{\rm 500c}} E(z)^{-4} \left( 1+z\right )^3.
\label{eq:corr}
\end{eqnarray}

From Eq.~\ref{eq:nhfromx} it follows that the recovered density scales as $\epsilon_c^{-1/2}$. The sensitivity of $\epsilon_c$ to the Galactic hydrogen column density, hot gas metallicity, and temperature are illustrated in Fig.\ref{fig:epsilon2}. The black curve ($Z/Z_\odot=0.2$) can be considered typical \citep[e.g.][]{2021A&A...646A..92G,2022A&A...663L...6A} for the outskirts of clusters  located far above the Galactic plane (hence low $N_H$). Corresponding variations of $\epsilon_c^{-1/2}$ with gas temperature are modest down to $kT \sim 0.3\,{\rm keV}$. For instance, the uncertainties in temperature at $3
\times R_{500c}$ (Fig.~\ref{fig:upp}) derived under the assumption of $\epsilon_c(T)={\rm const}$, imply $kT\approx 1\pm 0.3\,{\rm keV}$ . From Fig.\ref{fig:epsilon2} we can infer that corresponding variations of $\epsilon_c^{1/2}$ are $\sim 10$\%, i.e. a factor of $\sim 3$ smaller. The same is true for the uncertainty arising from the redshift dependence of $\epsilon_c$ (see Fig.~\ref{fig:epsilon}).
These arguments show that there is no internal inconsistency in the calculation of gas density and temperature arising from the assumption of $\epsilon_c(T)={\rm const}$. However, they do not provide direct proof that the X-ray-emitting gas is homogeneous in temperature and/or density. A more accurate answer would be possible with future microcalorimeter missions, especially those with full imaging capabilities like LEM \citep[][]{2022arXiv221109827K} that can measure both the mean temperature in the cluster outskirts as well as a distribution over gas temperatures using Fe-L complex and lines of lighter elements like Ne and O below 1~keV. In any case, X-ray observations place constraints only on the X-ray-emitting gas. Cooler gas, invisible in X-rays, might be present too. However, the broad agreement of the density profile derived in \S\ref{sec:res} with simulations suggests that the hot phase dominates at least up to $3\times R_{500c}$.

\section{Fiducial model parameters}
\label{app:bfm}
The multi-parameter model used to fit the X-ray surface brightness (see Eq.~\ref{eq:beta_mod}) is flexible enough to describe the data equally well with a slightly different combination of parameters. In this sense, only the overall shape of the recovered density distribution makes sense rather than the values of individual parameters. With this clause in mind, the following set of parameters yields the resulting density profile in the form 
%\begin{eqnarray}
%n_p(r)=N\times  \left [F(r,r_c,\alpha,\beta,r_s,\gamma,\epsilon,r_{s2},\lambda,\xi) \right ]^{1/2},
%\end{eqnarray}
%where $r$, $r_c$, $r_s$, $r_{s2}$ are units of $R_{\rm 500c}$. The best-fitting values of the %parameters are:
%$N=1.4\times 10^{-2}\,{\rm cm^{-3}}$, $r_c=0.187$, $\alpha=-0.7$, $\beta=0.193$, $r_s=20.2$, %$g=3.29$, $e=5.73$, $r_{s,2}=20.8$, $g2=0.698$, $e2=7.99$.  
\begin{eqnarray}
n_p(r)=N\times  \left [F(r,r_c,\alpha,\beta,r_s,\gamma,\epsilon,r_{s2},\lambda,\xi) \right ]^{1/2}, 
\end{eqnarray}
where $r$, $r_c$, $r_s$, $r_{s2}$ are in arcminutes and $R_{\rm 500c}=10$ arcmin by definition. The best-fitting values of the parameters are:
$N=1.4\times 10^{-2}\,{\rm cm^{-3}}$, $r_c=0.187$, $\alpha=-0.7$, $\beta=0.193$, $r_s=20.2$, $\gamma=3.29$, $\epsilon=5.73$, $r_{s,2}=20.8$, $\xi=0.698$, $\lambda=7.99$.

\begin{comment}
rc 1.86683E-01
b 1.93145E-01
a -6.99011E-01
rs 2.02300E+01
g 3.29097E+00
e 5.72767E+00
rs2 2.08131E+01
g2 6.98483E-01
e2 7.99005E+00
N 0.01406211012
\end{comment}

The relevant radial range, after exclusion of the outer regions where the signal is below 1\% of the background is $\sim (10^{-2}-3)\times R_{\rm 500c}$. 

For the entropy radial profile in the range $\sim (3\times 10^{-2}-3)\times R_{\rm 500c}$, the following power law fit provides reasonable approximation: 
\begin{eqnarray}
K_{\rm e}=1150 \left ({r/R_{\rm 500c}} \right )^{0.8}\,\,{\rm keV\,cm^2}.
\end{eqnarray}

\section{Wedge-to-wedge variations}
\label{app:wedges}

\begin{figure}
%\centering
\includegraphics[angle=0,trim=1cm 5cm 0cm 2cm,width=0.99\columnwidth]{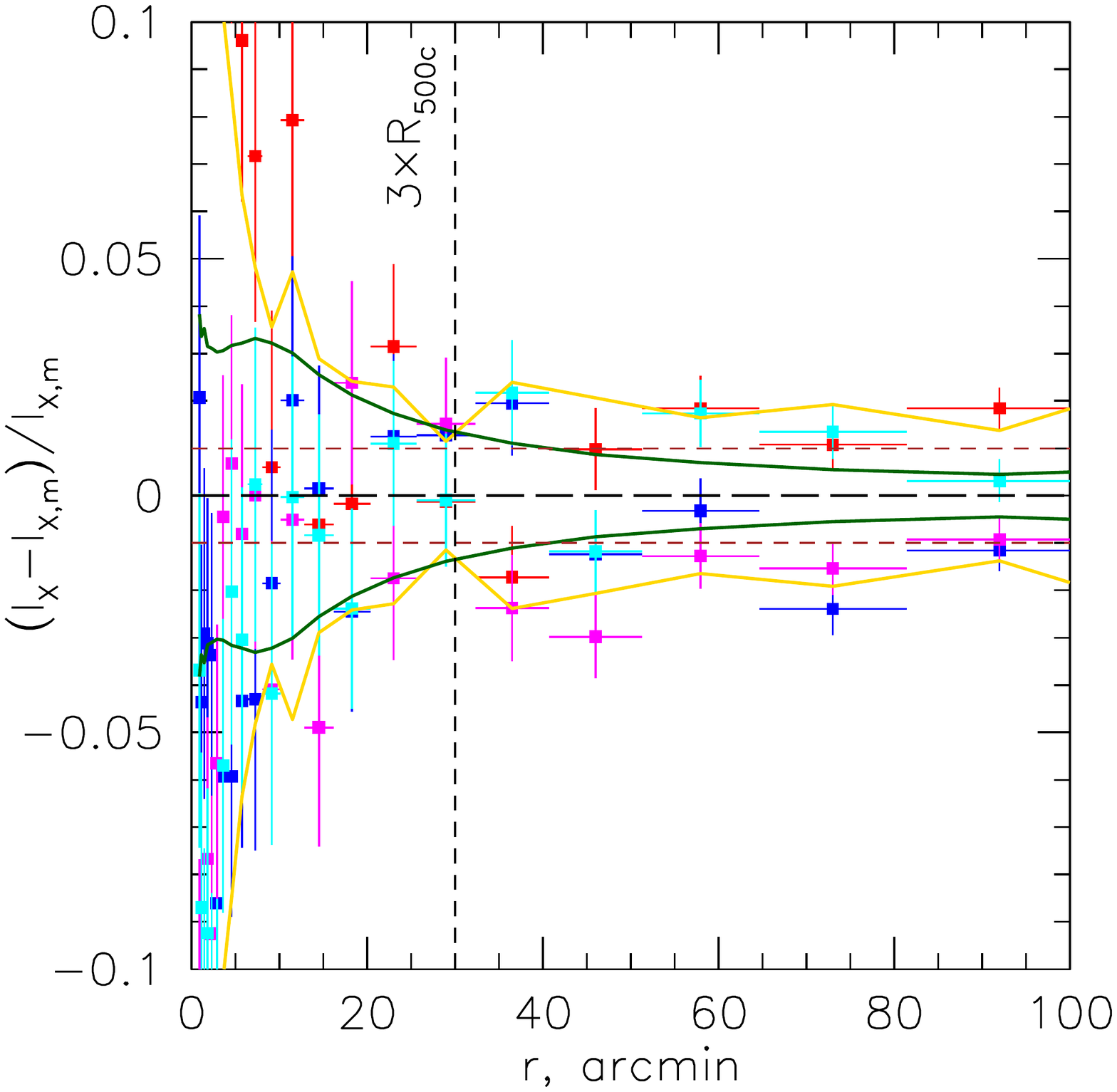}
\caption{Variations of the X-ray surfaces brightness (0.3-2.3~keV) in four $90^\circ$-wedges relative to the best fitting model $I_{\rm X,m}(r)$ to the azimuthally-averaged profile. The model includes both the cluster emission and a flat background. As in the previous plots, $10'$ corresponds to $R_{500c}$. The coloured points (red, blue, magenta, and cyan) correspond to different wedges.  The error bars include pure photon-counting noise and the estimated variance due to fluctuations in the number of unresolved X-ray sources. The two dark solid curves show  $\pm 1\sigma$ region, where $\sigma$ corresponds to errors assigned to individual measurements (photon counting noise and fluctuations caused by unresolved sources). The yellow curves show $\pm\,$r.m.s. of the deviations in four wedges evaluated for each radial bin. At very small radii (within $R_{500c}$), the relative variations are  $\sim$10\%, i.e. substantially larger than the statistical errors. In the most important region between $R_{500c}$ and $3\times R_{500c}$, the observed scatter matches expectations. Outside $3\times R_{500c}$ the scatter in the measured data points in wedges ($\sim2$\%) exceeds the assigned errors. If in the azimuthally-averaged profile (average of four wedges) these variations are independent, the expected level of noise will be factor 2 smaller, i.e. at the level of 1\% of the background. For comparison, the two horizontal lines show the level $\pm 1$\% from the background level.
} 
\label{fig:wedges}
\end{figure}

In \S\ref{sec:res}, an azimuthally-averaged X-ray surface brightness profile is discussed and approximated with an analytic function (see also Appendix~\ref{app:bfm}). Here we show (Fig.~\ref{fig:wedges}) deviations from this function in individual $90^\circ$-wedges. In this plot, $I_{X,m}$ stands for the full analytic model that includes both the cluster emission and a flat background. Inside the central region ($10'$ corresponds to $R_{500c}$), where the cluster emission dominates, deviations at the level of $\sim 10$\% are present, reflecting the residual variations of the surface brightness in the stacked image. These deviations might be caused by the uncertainties in the definition of cluster centres, intrinsic ellipticities of individual objects, etc. At large radii, the sky background (CXB and Milky Way foreground) dominates and the plot illustrates that the final scatter of the measured data points in wedges amounts to $\sim$2\% of the total background. 
%In the last two radial bins, the scatter appears to be somewhat larger than expected based on the estimated errors. A larger sample is needed to verify this statement. This will be done in a future study.

%%%%%%%%%%%%%%%%%%%%%%%%%%%%%%%%%%%%%%%%%%%%%%%%%%

% Don't change these lines
\bsp	% typesetting comment
\label{lastpage}
\end{document}